# Robust Zero-Field Skyrmion Formation in FeGe Epitaxial Thin Films

J. C. Gallagher[1], K. Y. Meng[1], J. T. Brangham[1], H. L. Wang[1], B. D. Esser[2], D. W. McComb[2], and F. Y. Yang[1]

[1]Department of Physics, The Ohio State University, Columbus, OH, 43210, USA

[2]Center for Electron Microscopy and Analysis, Department of Materials Science and Engineering, The Ohio State University, Columbus, OH, 43212, USA

Abstract

B20 phase magnetic materials, such as FeGe, have been of significant interests in recent years because they enable magnetic skyrmions, which can potentially lead to low energy cost spintronic applications. One major effort in this emerging field is the stabilization of skyrmions at room temperature and zero external magnetic field. We report the growth of phase-pure FeGe epitaxial thin films on Si(111) substrates by ultrahigh vacuum off-axis sputtering. The high crystalline quality of the FeGe films was confirmed by x-ray diffraction and scanning transmission electron microscopy. Hall effect measurements reveal strong topological Hall effect after subtracting out the ordinary and anomalous Hall effects, demonstrating the formation of high density skyrmions in FeGe films between 5 and 275 K. In particular, substantial topological Hall effect was observed at zero magnetic field, showing a robust skyrmion phase without the need of an external magnetic field.

Magnetic skyrmions are topological spin textures,[1–8] which have attracted significant interests in recent years due to their intriguing magnetic interactions and attractive attributes for magnetic storage and other spintronic applications.[3,4,9] The B20 phase FeGe, MnSi, MnGe and related compounds received particular attentions because in their non-centrosymmetric B20 cubic structure, the spin-orbit coupling induced Dzyaloshinskii-Moriya interaction (DMI) enables non-collinear spin alignment and skyrmion formation.[3,10,11] The topological spin texture of skyrmions allows skyrmion motion driven by an electrical current at a very low current density of $10^5$ to $10^6$ A/m$^2$, which is 4 to 5 orders of magnitude lower than those needed for domain wall motion in conventional ferromagnets.[12,13] For technological applications, it is desired to have skyrmion formation at room temperature and zero magnetic field. However, to date, B20 skyrmions can only be realized below room temperature and at finite magnetic fields.[1–9,12–16] A major effort in this emerging field is to increase their magnetic ordering temperature ($T_N$) and to stabilize the skyrmion phase at zero magnetic field. Thin films are a good avenue for achieving these goals since finite size effect can increase the range of temperatures ($T$) and magnetic fields ($H$) within which the skyrmion phase is stable.[4,16–18] FeGe has the highest $T_N$ of 278 K among the B20 skyrmion magnets.[3] Here, we report the observation of robust skyrmion formation at zero magnetic field in phase-pure FeGe epitaxial films.

FeGe epitaxial films were deposited by ultrahigh vacuum (UHV) off-axis sputtering.[19–24] To make the FeGe sputtering target, FeGe powder was synthesized by heating a mixture of Fe and Ge powders at 600°C in pure H$_2$ at atmospheric pressure for three hours, followed by two cycles of grinding and reheating in H$_2$. X-ray diffraction (XRD) scans reveal that the powder is a mixture of hexagonal B35 FeGe[25] and cubic B20 FeGe.[26] The FeGe powder was pressed into a 2" target and loaded into a UHV sputtering chamber with a base pressure of $4 \times 10^{-11}$ Torr. The ultraclean environment ensures the high purity of the deposited FeGe films. Si(111) substrates were first treated by HF immediately before loading into the sputter system. FeGe films were deposited on the Si(111) substrates at a temperature ($T$) of 290°C using DC sputtering with a constant current of 20 mA, which resulted in a deposition rate of 0.94 nm/minute (see Supplementary Materials for details on film growth at different temperatures). FeGe film thicknesses of 100, 65, and 36 nm are used for this study. All FeGe films exhibit pure B20 phase as shown in the XRD scans in Fig. 1a with no impurity phases detected. The out-of-plane lattice constants are 4.682 Å, 4.684 Å, and 4.685 Å for the 100, 65, and 36 nm FeGe films, respectively, which match well with Si with a 30° rotation (5.431 Å × cos 30° = 4.703 Å). Figure 1b shows the X-ray reflectivity (XRR) spectra for the 36 and 65 nm FeGe films. The clear XRR peaks indicate that the films are uniform, and from these peaks the thicknesses of the films were calibrated. Figure 1c shows the Φ-scans of the FeGe(200) and Si(400) peaks of the 100 nm film taken at a tilt angle of 57.4° from the film normal. The six-fold symmetry of the B20 FeGe(200) and the three-fold symmetry of the diamond-structure Si(400) have a 30° rotation relative to each other as expected from epitaxial growth.

The 100 nm FeGe film was imaged by Scanning Transmission Electron Microscopy (STEM) using an FEI probe-corrected Titan[3] 80-300 S/TEM. Figure 2a shows an STEM image of the FeGe/Si interface viewed along the $<110>_{FeGe}$ direction, which clearly reveals the B20 lattice of the FeGe film and the diamond structure of the Si substrate. There is a 1.2 to 1.5 nm interfacial region due to the formation of an Fe-Ge-Si interfacial layer in transition from diamond-Si to B20-FeGe. An atomic resolution STEM image in Fig. 2b shows two intertwined networks of parallelograms with a ~70° rotation between the two networks as illustrated in Fig. 2c. To the best of our knowledge, this is the first clear observation of the B20 atomic ordering of FeGe by STEM, demonstrating the high crystalline quality of the FeGe epitaxial films.

Out-of-plane magnetic hysteresis loops of the FeGe films were measured using a 7 tesla (T) Quantum Design Superconducting Quantum Interface Device (SQUID) magnetometer at temperatures ranging from 5 to 300 K. Figure 3a shows the hysteresis loop at $T = 5$ K for the 36 nm FeGe film, which exhibits the typical behavior for an out-of-plane magnetic field. The three FeGe films were patterned into a standard Hall bar structure with a width of 0.5 mm and a spacing of 2 mm between the two resistance measurement arms by optical lithography and $Ar^+$ ion etching. Longitudinal ($\rho_{xx}$) and Hall ($\rho_{xy}$) resistivity measurements were taken using a 14 T Quantum Design Physical Property Measurement System (PPMS). A constant current density of 2,000 A/cm$^2$ is applied using a Keithley 2400 current source meter while longitudinal and Hall voltages were measured using a Keithley 2182A nanovoltmeter to obtain $\rho_{xx}$ and $\rho_{xy}$. Figure 3b and its inset show the total Hall resistivity at 5 K for the 36 nm film, where we note three features: (1) a linear background at large fields (> 2 T), (2) a magnetic reversal behavior at intermediate fields that follows the magnetization hysteresis loop in Fig. 3a, and (3) a Hall hysteresis loop within ±3000 Oe that does not follow the magnetization hysteresis loop. These three features can be attributed to the ordinary Hall effect, anomalous Hall effect, and topological Hall (TH) effect, respectively. The total Hall resistivity is a combination of these three terms as defined by the equation:[14,15]

$$\rho_{xy} = R_o H + R_s M + \rho_{TH}, \quad (1)$$

where $R_o$ and $R_s$ are the ordinary and anomalous Hall coefficient, respectively, $M$ the out-of-plane magnetization, and $\rho_{TH}$ the topological Hall resistivity. When an electrical current is driven through the FeGe film, the electrons experience an emergent electromagnetic field through interaction with the skyrmions.[3] Consequently, the electrons are scattered off the skyrmions in a direction opposite of the anomalous Hall effect, generating a topological Hall voltage.[5,27,28] In Fig. 3b, the coercivity ($H_c$) of the Hall hysteresis loop is 2400 Oe, which is much larger than the $H_c =$ 240 Oe for the magnetization hysteresis (Fig. 3a); meanwhile, the Hall resistivity switches sign before reaching zero field. These are signatures of the topological Hall effect.[16]

The anomalous Hall coefficient can be modeled into a power-law form of $\rho_{xx}$, $R_s = b\rho_{xx}^2 + c\rho_{xx}$, where the quadratic $b\rho_{xx}^2$ term is due to a scattering independent mechanism and the linear $c\rho_{xx}$ term is caused by skew scattering.[27,29] A log-log plot of the anomalous Hall resistivity ($\rho_{AH}$) vs. $\rho_{xx}$ at $|H|$ = 4 T reveals a linear dependence with a slope of 2.3, suggesting that the anomalous Hall effect is dominated by the scattering independent mechanisms and the $c\rho_{xx}$ term can be neglected[29] (see Supplementary Materials for details). In addition, all FeGe films show very small magnetoresistance (<0.7% at fields up to 7 T for all samples); thus $R_s$ is approximately magnetic field independent. At $|H| \geq 2$ T, the FeGe films are in the saturated ferromagnetic state and the topological Hall effect is absent ($\rho_{TH}$ = 0) due to the lack of skyrmions. As a result, Eq. (1) can be simplified as,

$$\frac{\rho_{xy}}{H} = R_o + b\frac{\rho_{xx}^2 M_s}{H}. \quad (2)$$

By plotting $\frac{\rho_{xy}}{H}$ vs. $\frac{\rho_{xx}^2 M_s}{H}$, which exhibits a linear dependence, $R_o$ and $b$ can be obtained from the $y$-intercept and the slope, respectively.[14] The topological Hall resistivity was then extracted by subtracting out the ordinary and anomalous Hall resistivity in Eq. (1), as shown in Fig. 3c for the three FeGe epitaxial films (see Supplementary Materials for details on background subtraction). Three notable features are observed in Fig. 3c. First, strong topological Hall effect ($\rho_{TH}$ = 94.8 ± 1.6 nΩ-cm for the 36 nm film) is detected at 5 K, which is significantly lower than the temperature window for skyrmion formation in bulk FeGe.[5,30] Second, all three films exhibit clear hysteresis where the switching of $\rho_{TH}$ is opposite to those of the anomalous Hall effect and magnetization hysteresis (indicated by the arrows). Third, there are significant remanent values of $\rho_{TH}$ at $H = 0$ for all three films, which demonstrates robust skyrmion formation at zero field. This stable skyrmion phase without the need of an applied magnetic field is desired for technological applications.

As the temperature increases, $\rho_{TH}$ continues to increase up to 275 K, indicating the enhancement of the topological Hall effect. Figure 3d shows the topological Hall resistivities at $T$ = 250 K for the three FeGe films. The topological Hall effect is very large for all three FeGe films, e.g., $\rho_{TH}$ = 918 ± 5 nΩ-cm for the 36 nm film, which is more than five times larger than the highest values of $\rho_{TH}$ previously reported for B20 skyrmion materials.[4,13,15] Figures 4a-4f show the field (-2 to 2 T) and temperature (5 to 275 K) dependencies of $\rho_{TH}$ for the 100, 65, and 36 nm films, where both the decreasing (left column) and increasing (right column) field branches of the hysteresis loops are presented. Figure 4 shows that the range of skyrmion phase and the magnitude of $\rho_{TH}$ depend on the FeGe thickness. In bulk FeGe, the skyrmion phase is metastable, with its stable counterpart being a helical phase with a helical period of 68-70 nm as measured by neutron scattering.[30] It has been shown that by using geometric constraints, the helical phase is less stable and the skyrmion phase becomes more stable.[4,5,14,16] For the 100 nm film, the thickness is about 1.5 times larger than the helical period observed in bulk FeGe, while its skyrmion phase is much broader and stronger than in the bulk. For the 65 nm film (Figs. 4c and 4d), of which the thickness is just below the helical period of the bulk, the skyrmion phase is stabilized in a broader field range and becomes stronger as compared with that of the 100 nm film at low temperatures. At 36 nm, the film thickness is about half the helical period of the bulk FeGe (Figs. 4e and 4f) and the skyrmion phase expands further in both the field and temperature axes. Stabilization of the skyrmion phase in epitaxial FeGe films likely arises from the epitaxy with the Si substrate and small thicknesses, which suppress the helical phase and favor skyrmion formation.

Figure 5a shows the magnetic field at the maximum $\rho_{TH}$ for the three FeGe films at all temperatures. It indicates that with decreasing thickness the skyrmion phase shifts toward lower fields, which means skyrmion formation requires lower energy for thinner films. The lower energy costs for skyrmion formation at smaller thicknesses also result in higher overall skyrmion densities, thus larger $\rho_{TH}$, as shown in Fig. 5b.

The main result of this work is the robust skyrmion formation at zero magnetic field, as can be seen from the large remanent values of $\rho_{TH}$ in Fig. 3c. Figures 4g and 4h show a zoom-in view of $\rho_{TH}$ for the 36 nm film (-0.15 to 0.15 T) to highlight the hysteretic nature and the large remanent $\rho_{TH}$ between 5 and 275 K. The large remanent $\rho_{TH}$ indicates that a significantly high skyrmion density persists at zero magnetic field. To quantitatively characterize the zero-field skyrmion stability, we define the *Squareness* of $\rho_{TH}$, *Squareness* $\equiv \rho_{TH}(H = 0)/\rho_{TH}(\max)$, and plot it as a function of temperature for the three FeGe films. The *Squareness* is as high as 0.77 for the 100 nm film at $T = 5$ K and decreases with increasing temperature. Substantial *Squareness* persists to high temperatures, especially for the thinner films, for example, *Squareness* = 0.39 at 150 K for the 36 nm film.

In summary, the robust skyrmion formation at zero field as revealed by the large remanent topological Hall resistivity in FeGe films demonstrates that skyrmions can be stabilized as the ground state (zero field) in epitaxial thin films, which is a major step forward in the development of skyrmion spintronic applications. This is enabled by the deposition of pure B20 phase, high quality epitaxial FeGe thin films using off-axis UHV sputtering. The large topological Hall resistivity (up to $1067 \pm 3$ n$\Omega$-cm) observed in the FeGe films is much larger than previously reported, indicating a high density of skyrmions which can be created at a low energy cost.

**Acknowledgements**

This work was primarily supported by the U.S. Department of Energy (DOE), Office of Science, Basic Energy Sciences, under Grants No. DE-SC0001304 (sample growth and characterizations of structural and transport properties). This work was supported in part by the National Science Foundation under Grant No. DMR-1507274 (magnetization measurements) and Center for Emergent Materials, an NSF-funded MRSEC, under Grant No. DMR-1420451 (STEM characterization).

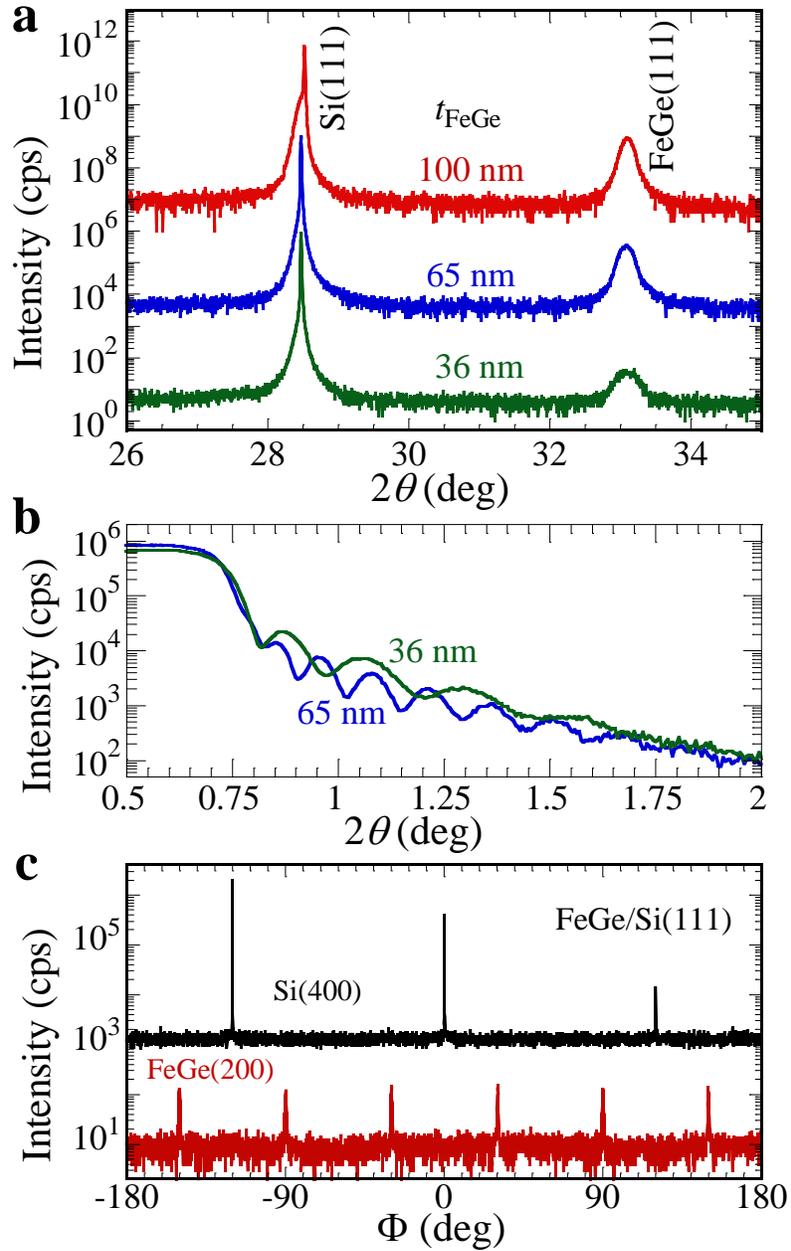

**Figure 1: a** Semi-log $2\theta$-$\omega$ XRD scans of 100 nm (red), 65 nm (blue), and 36 nm (green) FeGe films epitaxially grown on Si(111). **b** XRR scans of the 65 nm (blue) and 36 nm (green) FeGe films. **c** $\Phi$-scans of the Si(400) peaks (black) and FeGe(200) peaks (red) of the 100 nm FeGe reveal a 30° rotation between the Si and FeGe cubic lattices.

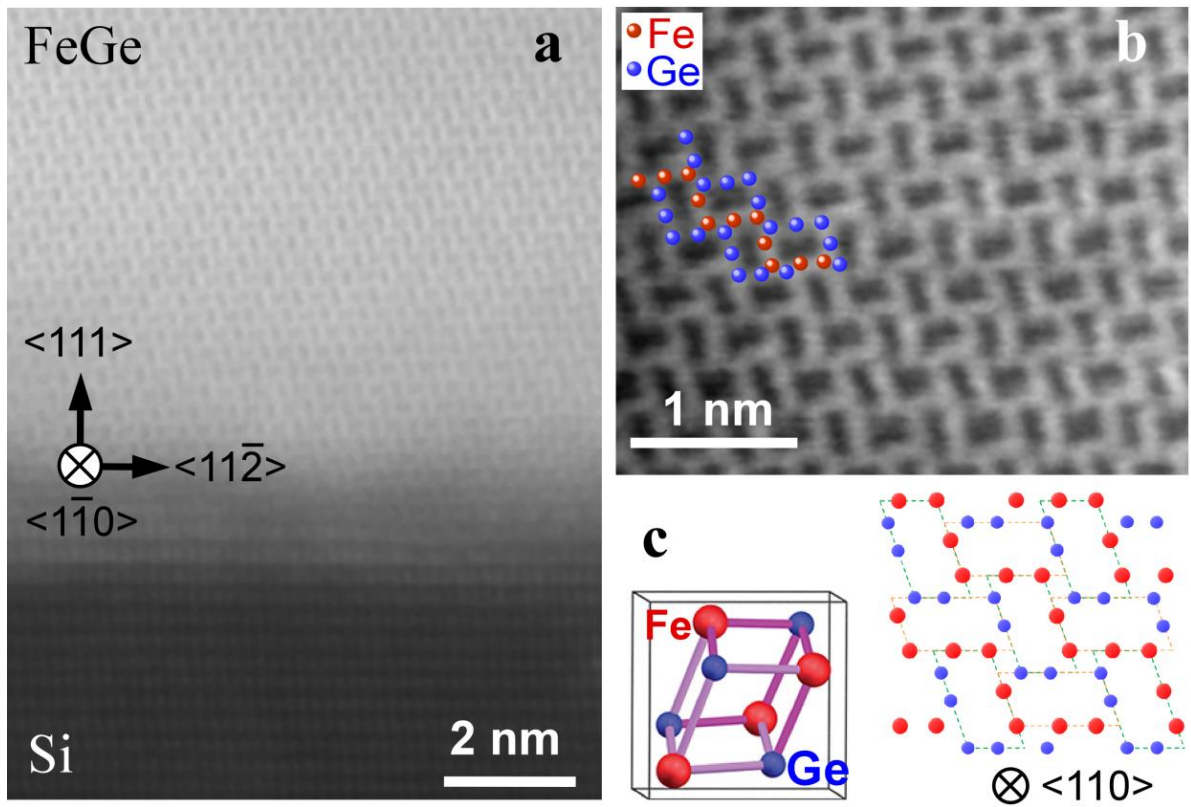

**Figure 2**: **a** STEM image of a 100 nm FeGe thin film viewed along the FeGe <110> axis near the interface with Si(111). A 1.2 - 1.5 nm interfacial region is detected, which arises from the transition from the diamond structure of Si to the B20 structure of FeGe. **b** A high resolution STEM image of the FeGe film reveals the B20 ordering of Fe and Ge atoms. **c** Schematic of the cubic lattice and the <110> projection of the B20 structure.

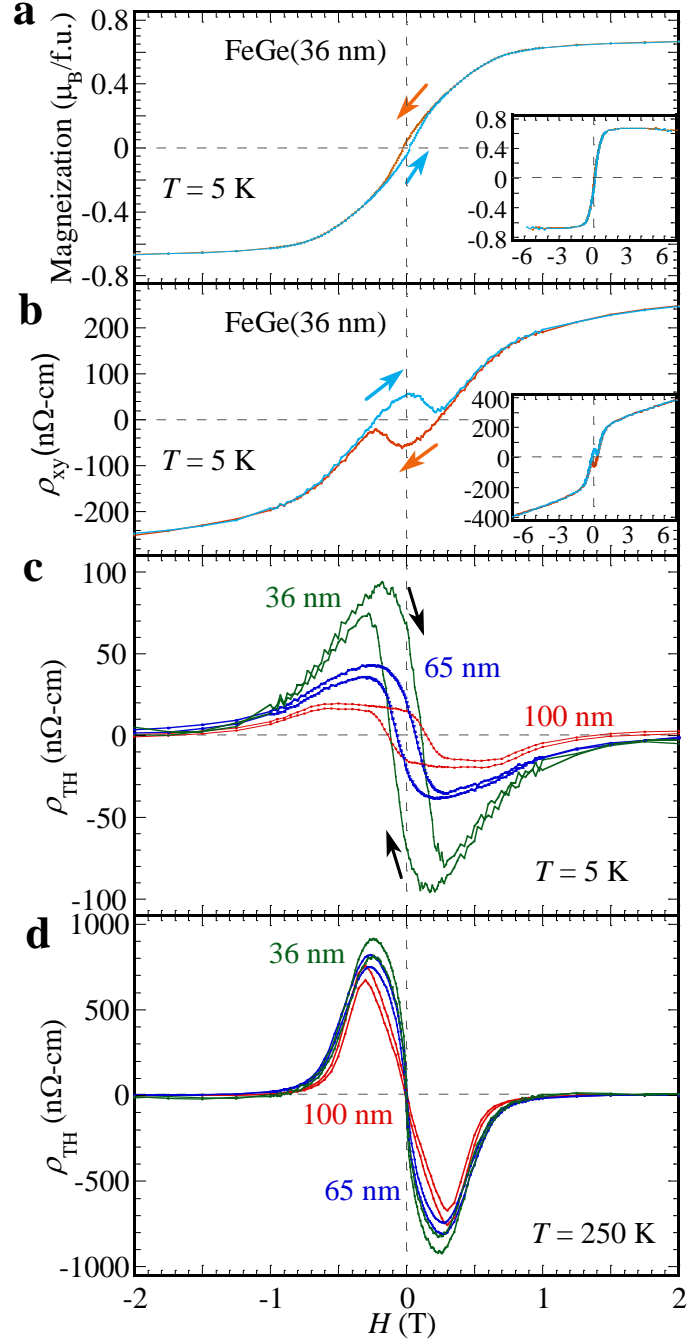

**Figure 3**: **a** Out-of-plane magnetic hysteresis loop of the 36 nm FeGe film taken at 5 K shows a small coercivity of 500 Oe after subtraction of the diamagnetic background from Si. The arrows indicate sweeping directions of the magnetic field. **b** Total Hall resistivity ($\rho_{xy}$) of the 36 nm FeGe film with a coercivity of 2400 Oe ($T = 5$ K) reveals a dominant topological Hall effect, which has a path opposite of the expected anomalous Hall effect. The insets in **a** and **b** show the full hysteresis loops between -7 and +7 T. The topological Hall resistivity ($\rho_{TH}$) hysteresis loops for the 36 nm (green), 65 nm (blue), and 100 nm (red) FeGe films at **c** 5 K and **d** 250 K show a clear skyrmion phase. At $T = 5$ K, $\rho_{TH}$ exhibits substantial remanent values at $H = 0$, demonstrating robust skyrmion formation in the absence of magnetic field.

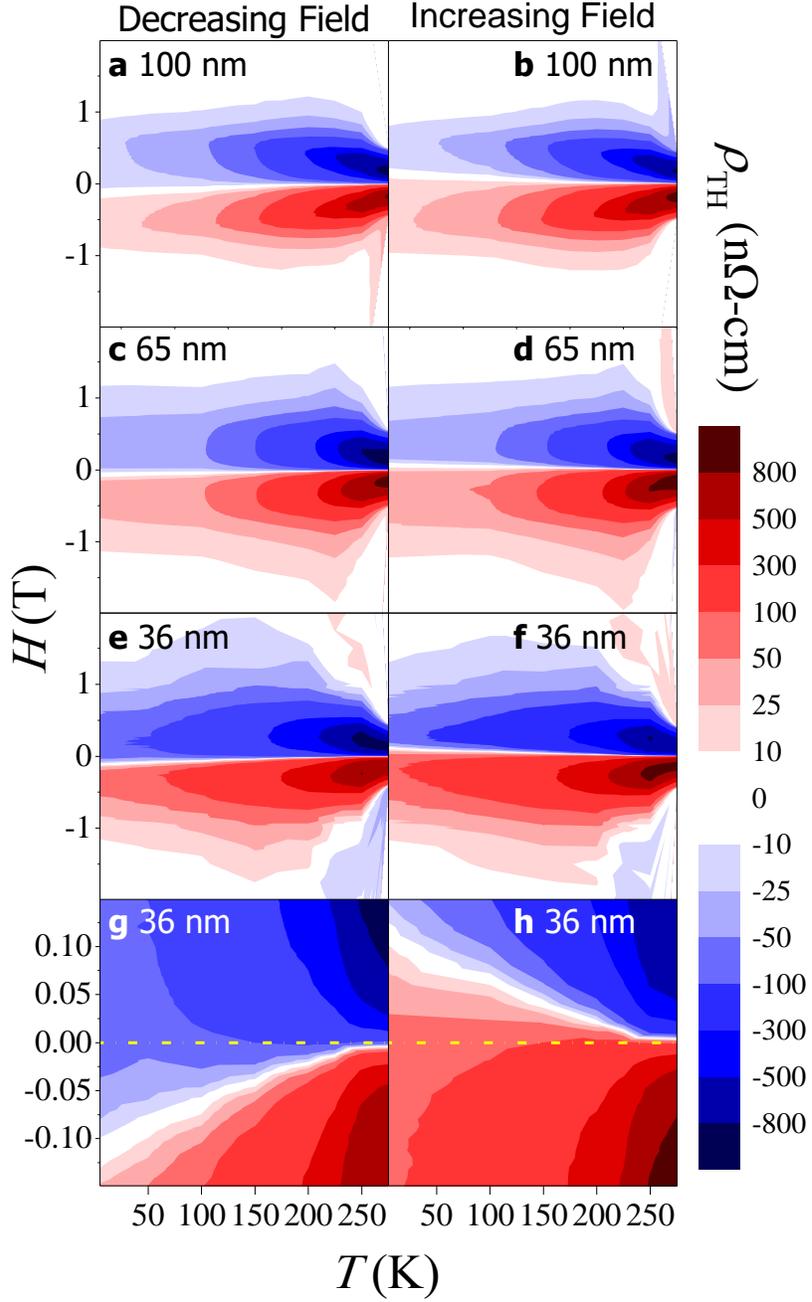

**Figure 4**: Contour plots of $\rho_{TH}$ for the **a, b** 100 nm, **c, d** 65 nm, and **e-h** 36 nm FeGe films. The plots on the left have the field sweeping from +7 to -7 T and the plots on the right have the field sweeping in the opposite direction. A zoom-in view of the 36 nm film in **g** and **h** highlights that nonzero topological Hall resistivity exists at zero field for all temperatures, particularly at low temperatures with substantial remanence at $H = 0$.

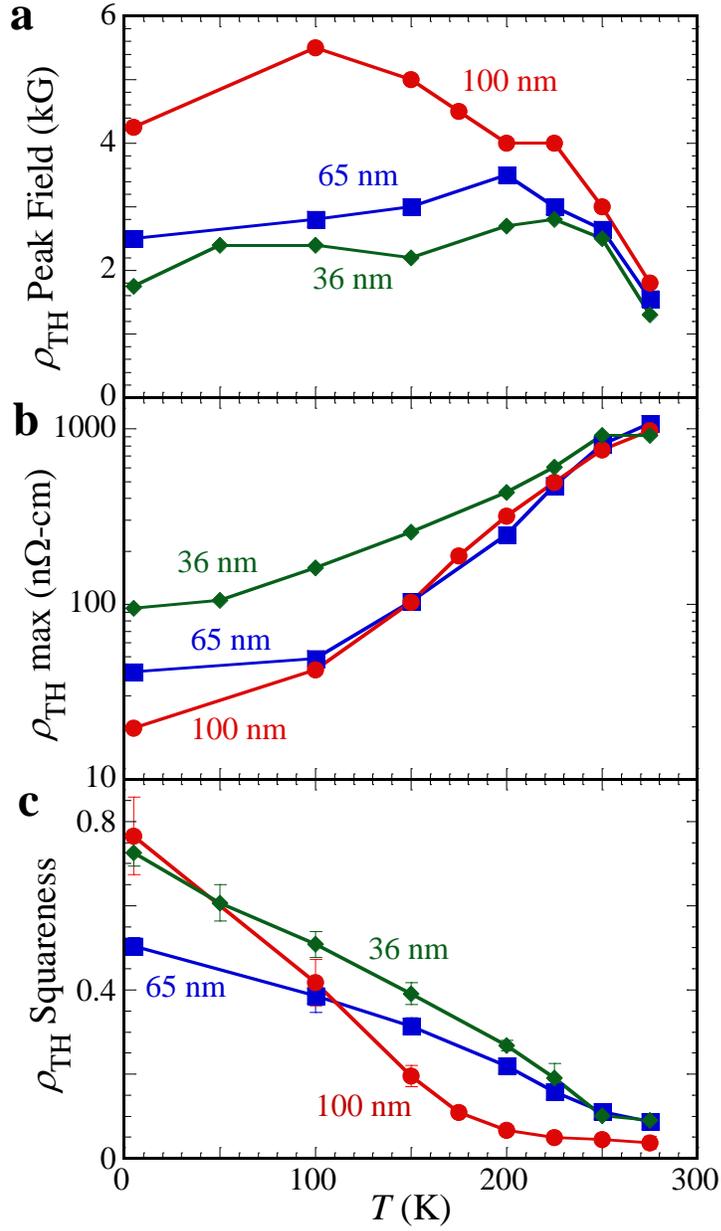

**Figure 5:** **a** Temperature dependence of the magnetic field at the maximum topological Hall resistivity for the 36 (green), 65 (blue), and 100 nm (red) FeGe films. The field for maximum skyrmion stability decreases at smaller film thickness, indicating that thinner films are desired for stabilizing skyrmion phases at lower fields. **b** Semi-log plot of the temperature dependence of the maximum $\rho_{TH}$ for the three samples. **c** The *Squareness* of the topological Hall resistivity, $\rho_{TH}(H = 0)/\rho_{TH}(\max)$, for the three FeGe films between 5 and 275 K, which reflects the stability of skyrmions at zero field.

## Supplementary Material

**Optimization of FeGe Growth Conditions**

FeGe epitaxial films were grown on Si(111) substrates using ultrahigh vacuum off-axis magnetron sputtering at various substrate temperatures ($T_s$) to identify the optimal deposition temperatures. Figure S1 below shows the $2\theta$-$\omega$ x-ray diffraction scans of four FeGe films grown at $T_s$ = 420°C, 315°C, 290°C, and 265°C. The deposition temperature of 290°C results in only one film peak at $2\theta \approx 33°$, which is the expected position of the FeGe(111) peak of B20 phase, demonstrating the growth of phase-pure B20-FeGe(111) epitaxial film on Si(111). Figure S2 shows the full range XRD scan between 10° and 80° where no impurity peaks are observed. At higher substrate temperatures (Figs. S1a and S1b), the Ge(111) peak at $2\theta \approx 27.3°$ appears, indicating the formation of Ge epitaxial phase. At lower substrate temperatures (Fig. S1d), an impurity peak at $2\theta \sim 36°$ appears, which can be attributed to the formation of hexagonal B35-phase FeGe [see reference 25 in main text]. The optimal temperature window for the epitaxial deposition of pure B20-phase FeGe films is between 280°C and 300°C.

**FeGe Topological Hall Effect Extraction**

The anomalous Hall (AH) resistivity ($\rho_{AH}$) can be modeled into a power law form of longitudinal resistivity ($\rho_{xx}$),

$$\rho_{AH} = R_s M = (b\rho_{xx}^2 + c\rho_{xx})M,$$

where the quadratic term ($b\rho_{xx}^2$) is due to a scattering independent mechanism and the linear term ($c\rho_{xx}$) is caused by skew scattering, both dependent on $\rho_{xx}$. All of our FeGe films show very small magnetoresistance (<0.7% at fields up to 7 T) as shown in Fig. S3a, indicating that $R_s$ is approximately magnetic field independent. A log-log plot of $\rho_{AH}$ vs. $\rho_{xx}$ at $|H| = 4$ T (where the magnetization is saturated) and at various temperatures reveals a linear dependence with a slope of 2.3 (Fig. S3b), suggesting that the anomalous Hall effect is dominated by the scattering independent mechanisms and the $c\rho_{xx}$ term can be neglected [see references 14, 27, and 29 in the main text]. At $|H| \geq 2$ T, the FeGe films are in the saturated ferromagnetic state and the topological Hall effect is absent ($\rho_{TH} = 0$) due to the lack of skyrmions. Given that $\frac{\rho_{xy}}{H} = R_o + b\frac{\rho_{xx}^2 M_s}{H}$ [Eq. (2) in main text], by plotting $\frac{\rho_{xy}}{H}$ vs. $\frac{\rho_{xx}^2 M_s}{H}$ at $|H| \geq 2$ T, which exhibits a linear dependence as shown in Fig. S4, $R_o$ and $b$ can be obtained from the y-intercept and the slope, respectively.

The procedure to extract topological Hall resistivity is shown in Fig. S5. Figure S5a shows the field dependence of total Hall resistivity. At high fields ($H \geq 2$) the data are linear and its slope represents the ordinary Hall coefficient $R_o$. After subtraction of the ordinary Hall effect, Fig. S5b looks similar to but does not follow the magnetization hysteresis loop in Fig. 3a (see main text). Using the extracted parameter $b$ and magnetoresistivity data (Fig. S3a), we subtracted the anomalous Hall effect contribution and obtained the topological Hall resistivity as shown in Fig S5c.

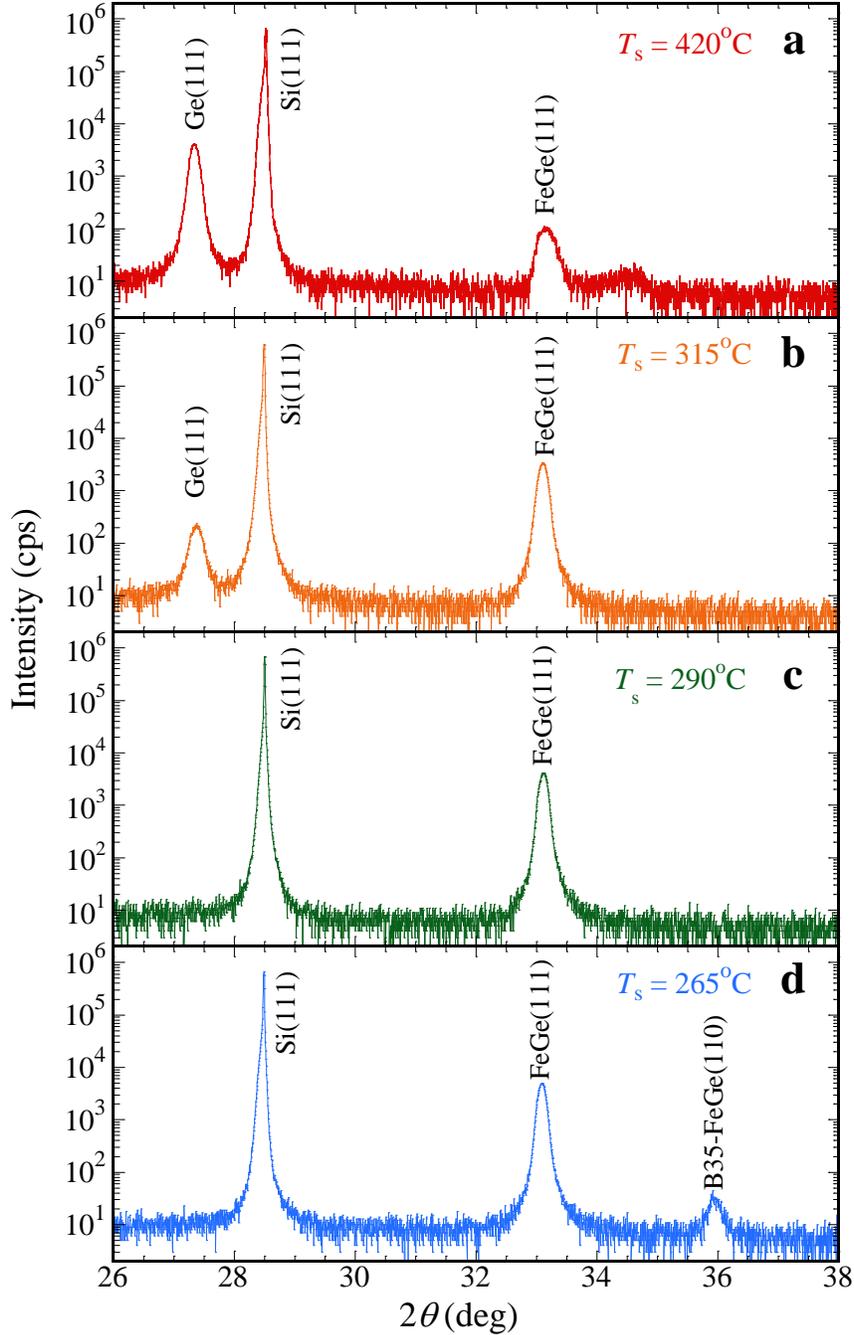

**Figure S1:** Semi-log $2\theta$-$\omega$ XRD scans for 200-nm FeGe films grown on Si(111) at substrate temperatures of **a** 420°C, **b** 315°C, **c** 290°C, and **d** 265°C. The XRD results indicate that 290°C is the optimal deposition temperature which gives a pure B20 phase, (111)-oriented epitaxial FeGe film on Si(111). Higher substrate temperatures lead to the growth of Ge(111) phase while at lower temperatures, an impurity peak at $2\theta \sim 36°$ appears, which is likely due to the formation of hexagonal B35-phase FeGe. The optimal temperature window for B20-phase FeGe epitaxial growth is rather narrow, between 280°C and 300°C.

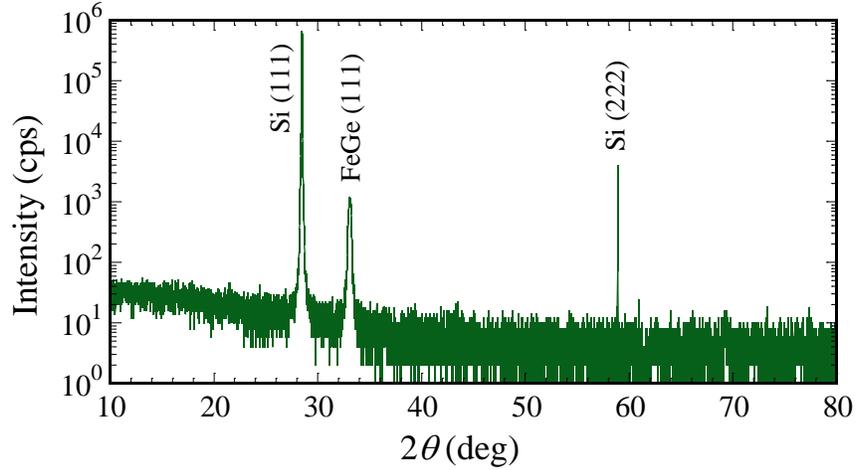

**Figure S2:** Full range semi-log $2\theta$-$\omega$ XRD scan for a 200-nm FeGe film grown on Si(111) at substrate temperature of 290°C, demonstrating pure B20 phase of the FeGe epitaxial film.

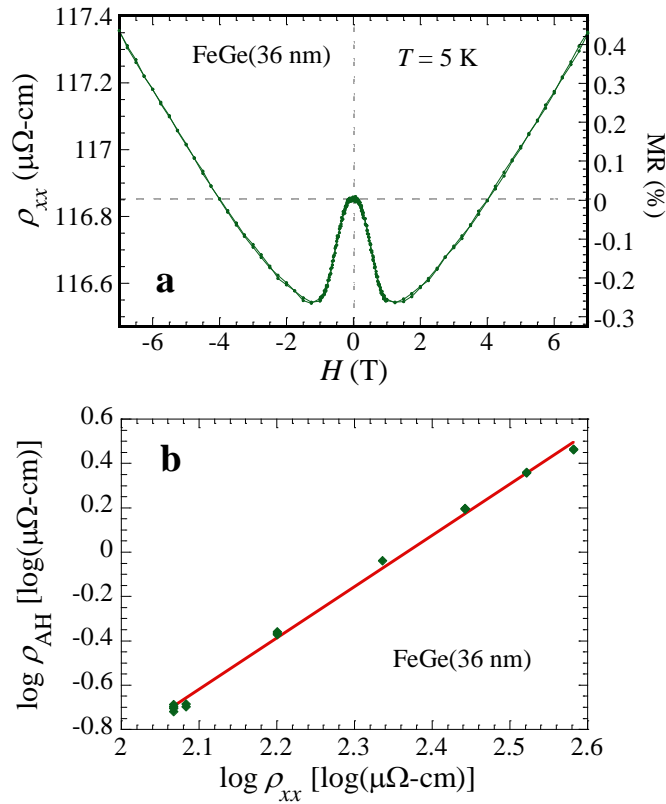

**Figure S3: a** Magnetoresistance of a 36 nm FeGe film at 5 K, which is less than 0.7%. **b** A log-log plot of the anomalous Hall resistivity ($\rho_{AH}$) vs the longitudinal resistivity ($\rho_{xx}$) taken at $|H| = 4$ T and various temperatures. A least-squares fit (red) reveals a slope of 2.3.

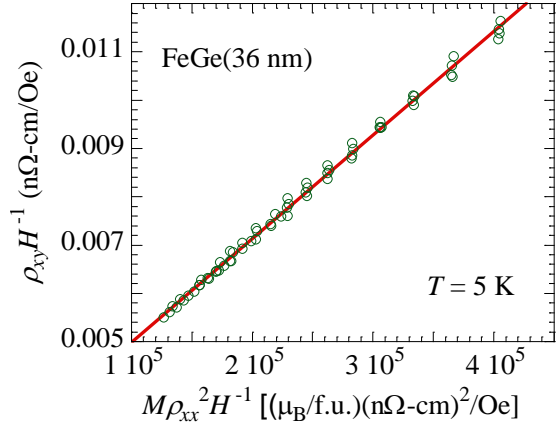

**Figure S4:** Plot of $\frac{\rho_{xy}}{H}$ vs. $\frac{\rho_{xx}^2 M_s}{H}$ yields a linear relationship. The slope and the y-intercept of the linear fit are coefficients $b$ and $R_o$, respectively.

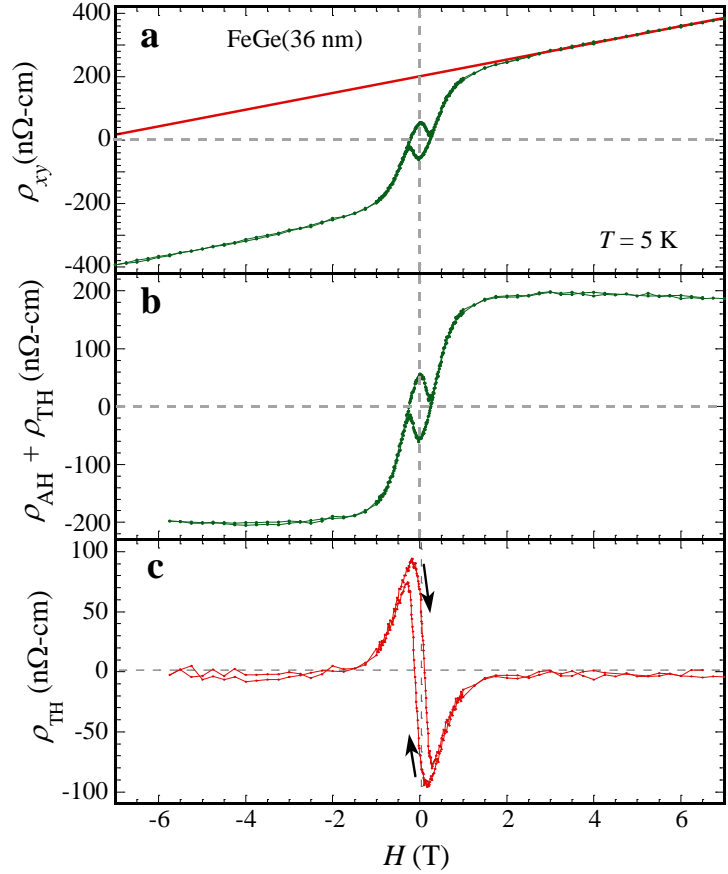

**Figure S5. a** The total Hall resistivity, $\rho_{xy}$ (green curve) and high field linear fit (red) for a 36 nm FeGe film taken at 5 K. The ordinary Hall effect can be extracted from the slope of linear fit at high field or the method described in Fig. S4. **b** The Hall resistivity after subtracting the ordinary Hall effect. **c** The topological Hall resistivity after further subtraction of the anomalous Hall effect by using the coefficient $b$, $\rho_{xx}$ and magnetization hysteresis obtained from Fig. S4, Fig. S3a, and Fig. 1a (main text), respectively.